\documentclass[twocolumn]{IEEEtran} 

\usepackage[pdftex]{graphicx}% Include figure files
\usepackage{dcolumn}% Align table columns on decimal point
\usepackage{bm}% bold math

\begin{document}
%\preprint{October 2006}

\title{$Q$-based design equations for resonant metamaterials and experimental validation}% Force line breaks with \\

\author{
Steven A. Cummer, \IEEEmembership{Senior Member, IEEE,} Bogdan-Ioan Popa, and Thomas H. Hand
\thanks{Manuscript received X.}
\thanks{This work was supported by DARPA through Contract No. HR001Ð05ÐC-0068. }
\thanks{S. A. Cummer, B.-I. Popa, and T. H. Hand are with the Electrical and Computer Engineering Department, Duke University, Durham, NC, USA.  E-mail: cummer@ee.duke.edu.}
}

\markboth{Submitted to IEEE Transactions on Antennas and Propagation}{Cummer {\it et al.}: Metamaterial Design Equations}

\maketitle

\begin{abstract}
Practical design parameters of resonant metamaterials, such as loss tangent, are derived in terms of the quality factor $Q$ of the resonant effective medium permeability or permittivity.  Through electromagnetic simulations of loop-based resonant particles, it is also shown that the $Q$ of the effective medium response is essentially equal to the $Q$ of an individual resonant particle.  Thus, by measuring the $Q$ of a single fabricated metamaterial particle, the effective permeability or permittivity of a metamaterial can be calculated simply and accurately without requiring complex simulations, fabrication, or measurements.  Experimental validation shows that the complex permeability analytically estimated from the measured $Q$ of a single fabricated self-resonant loop agrees with the complex permeability extracted from $S$ parameter measurements of a metamaterial slab  to better than 20\%.  This $Q$ equivalence reduces the design of a metamaterial to meet a given loss constraint to the simpler problem of the design of a resonant particle to meet a specific $Q$ constraint.  This analysis also yields simple analytical expressions for estimating the loss tangent of a planar loop magnetic metamaterial due to ohmic losses.  It is shown that $\tan \delta \approx 0.001$ is a strong lower bound for magnetic loss tangents for frequencies not too far from 1 GHz.  The ohmic loss of the metamaterial varies inversely with the electrical size of the metamaterial particle, indicating that there is a loss penalty for reducing the particle size at a fixed frequency.  
\end{abstract}

\section{Introduction}

\PARstart{A}{nalytical} calculations have provided important insight into how the effective electromagnetic parameters of engineered electromagnetic metamaterial structures behave \cite{pendry99,marques02}.  But accurate, quantitative determination of those parameters has been demonstrated primarily through complex simulations or experimental measurements.  Electromagnetic simulations \cite[and many others]{smith02a} and experimental measurements \cite{starr04} of the reflection and transmission coefficients of a metamaterial slab have been used to extract the effective permittivity and permeability through a well-established approach \cite{chen04ret,smith05ret}.  Different experimental measurements, including prism refraction experiments \cite{parazzoli03,huangfu04prism} and measurements of spatial field distributions inside the metamaterial \cite{cummer04nim,popa05intfields}, have also been used to infer the effective electromagnetic parameters of metamaterials.  These experimental procedures usually require samples of a specific size and shape, either to fill the entire cross sectional area of a waveguide or to be large enough in a free space measurement that edge effects are not important.  A procedure through which metamaterial parameters could be accurately estimated either analytically or from measurements of small material samples would enable simpler metamaterial design.  

Many metamaterials are based on magnetically or electrically self-resonant particles so that a wide range of effective parameters can be obtained in a controlled manner.  There are also universal relationships between resonance parameters that apply to almost all resonant phenomena.  Motivated by this, we show how, by experimentally measuring or analytically estimating the quality factor $Q$ of the underlying resonant particle, the frequency dependent effective parameters of a resonant metamaterial can be accurately estimated.  Simple analytical forms for important specific parameters, such as the effective loss tangent, in terms of $Q$ and the geometry of the unit cell are also derived.  Although effective metamaterial parameters can be written in terms of equivalent circuit parameters \cite{pendry99,popa06srrtheory}, in practice it can be difficult to compute the effective resistance of a structure because of spatially nonuniform current flow in the conductors.  Losses in any dielectric substrate are a function of the complicated field distribution of the particle near resonance and are difficult to predict without detailed simulations.  $Q$ measurements or simulations, in contrast, include all loss mechanisms automatically.  The overall approach is validated through experimental measurements.  We specifically analyze loop-based resonators for magnetic metamaterials, but the approach is applicable to any resonant-type metamaterial.

\section{Quality Factor Equivalence}

Two distinct resonant responses are involved in a metamaterial, each with its own quality factor $Q$.  An individual resonant particle has a $Q$, which we call $Q_{\rm part}$.  The bulk magnetic (i.e., permeability) or electric (i.e., permittivity) material response of a metamaterial composed of these resonant particles also has a resonant frequency dependence \cite{pendry99}, which we call $Q_{\rm mat}$.  

We assert that under conditions relevant to metamaterials, these two quality factors are essentially equal.  Resonator coupling can in general alter the $Q$ of individual resonances, and this does occur when the individual particles are assembled into a metamaterial array.  But metamaterials are typically designed so that the coupled resonant particles all collectively resonate at a frequency close to the resonance of an individual particle.  Losses are not shifted from one resonance to another, nor are additional loss mechanisms created in an assembly of these particles.  

\begin{figure}
\includegraphics[width=3.2in]{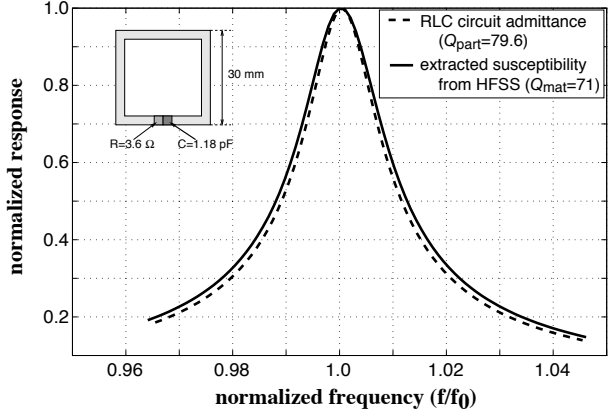}
\caption{Comparison of normalized resonant responses of the analytically computed series impedance of an individual loop particle (geometry in inset) and of the numerically computed effective magnetic susceptibility of an assembly of these loop particles.}
\label{fig:qsim}
\end{figure}

We demonstrate this $Q$ equivalence through electromagnetic simulations using the well-known loop resonator \cite{pendry99}.  We designed a capacitively loaded loop with a lumped resistor in series with otherwise lossless metal.  The inset in Figure \ref{fig:qsim} shows the specific element values ($R=3.58$ $\Omega$ and $C=1.18$ pF) and geometry of the 30 mm loop.  The circuit parameter extraction code Ansoft Q3D was used to extract an inductance of 95.7 nH for this loop.  Assuming the particle is sufficiently small, it is an RLC circuit with $Q_{\rm part}=(\sqrt{L/C})/R=79.6$ and a self-resonant frequency of $1/\sqrt{LC}=474$ MHz.  The loop is consequently 21 times smaller than a wavelength at resonance and the circuit approximation is justified.

The effective magnetic material response (i.e., the permeability) of a collection of these individual resonators also exhibits resonant behavior defined by (the $\exp(+j\omega t)$ sign convention is assumed throughout) 
\begin{equation}
\label{eq:mu}
\mu=1+\chi_m=1+\frac{F \omega^2}{(\omega_0^2-\omega^2+j\omega\omega_0/Q_{\rm mat})},
\end{equation}
where $\mu$ is the relative permeability of the medium and $\omega_0$ is the resonant frequency of the effective magnetic response.  The loss term is written in terms of $Q_{\rm mat}$, which is the quality factor of the resonant-form magnetic susceptibility $\chi_m=\mu-1$.  Analytical theory \cite{pendry99,popa06srrtheory} and simulations \cite{smith02a} have all shown that the magnetic response of such a material follows this functional form.  Although this form is not exactly a constant-numerator Lorentzian, the difference is not critical because the frequency range of interest is generally in a narrow band around the resonant frequency and thus $\omega\approx\omega_0$.  Nevertheless, the Lorentzian form is useful in a number of ways---for example, one can compute $Q$ graphically from a Lorentzian response from the maximum magnitude divided by the full-width half power bandwidth.  To obtain a pure Lorentzian, we can simply normalize the susceptibility to $\chi_m/\omega^2$.

We used Ansoft HFSS to compute the $S$ parameters of this loop structure placed in a cubic unit cell 40 mm on a side, and the effective permeability was extracted using the standard approach \cite{smith05ret}.  Figure \ref{fig:qsim} overlays the resonant $|\chi_m|/\omega^2$ and the series RLC admittance of a single particle based on circuit theory (both are amplitude normalized to unity).  The frequency dependence of these two frequency responses are in close agreement, with $Q_{\rm part}=79.6$ and $Q_{\rm mat}=71$.  Other simulations with higher and lower $Q$s also revealed the same basic $Q$ equivalence with 10\% or better agreement. This functional form for $\mu$ breaks down when the particles are not sufficiently small compared to wavelength, at which point spatial dispersion becomes important \cite{koschny03antires} and loop radiation resistance cannot be completely neglected.  However, under most practically useful metamaterial conditions, $Q_{\rm part}\approx Q_{\rm mat}$.

The value of this equivalence is that $Q_{\rm part}$ can be measured from a single fabricated element or from a small material sample that may not be suitable for a full reflection/transmission measurement.  $Q_{\rm part}$ can also be analytically estimated under many conditions.  This $Q$ equivalence thus enables rapid design analysis without complex simulations and experiments.

The other parameter in (\ref{eq:mu}), $F$, also plays a role in the achievable values of $\mu$.  It can be shown analytically that, for planar loop resonator configurations, $F=\mu_0 A_{\rm loop}^2/(V_{\rm cell}L)$ \cite{popa06srrtheory}, where $A_{\rm loop}$ is the area enclosed by a single loop resonator, $V_{\rm cell}$ is the volume of a unit cell, and $L$ is the inductance of a single loop resonator.  For planar loop resonator configurations, $F$ typically falls in the range $0.16-0.50$, depending on how densely packed the individual resonators are \cite{popa06srrtheory}.  We use this numerical range in the analysis below.

\section{$Q$-Based Metamaterial Design Equations}

With $F$ numerically bounded, $Q_{\rm mat}$ dominates the achievable effective properties of a metamaterial.  The $Q$ equivalence demonstrated above shows in turn that these properties are completely dictated by $Q_{\rm part}$.  Consequently designing a metamaterial to achieve a specific objective (loss tangent, minimum negative permeability or permittivity, etc) is reduced to the simpler problem of designing and fabricating a resonant particle with a given $Q$.  

We now rearrange (\ref{eq:mu}) to provide explicit equations for key metamaterial properties in terms of $Q$ (no subscript since $Q_{\rm part}$ and $Q_{\rm mat}$ are essentially equal) and $F$.  Defining $\chi_r={\rm Re}(\chi_m)$ and $\chi_i={\rm Im}(\chi_m)$, it is straightforward to show that
\begin{eqnarray}
|\chi_r|=\frac{F \omega^2 |\omega_0^2-\omega^2|}{(\omega_0^2-\omega^2)^2+\omega_0^2 \omega^2/Q^2} \label{eq:chir1}\\
|\chi_i|=\frac{F \omega_0 \omega^3/Q}{(\omega_0^2-\omega^2)^2+\omega_0^2 \omega^2/Q^2} \label{eq:chii1}\\
|\chi_m|^2=\frac{F^2 \omega^4}{(\omega_0^2-\omega^2)^2+\omega_0^2 \omega^2/Q^2}.
\end{eqnarray}
The ratio $|\chi_i|/|\chi_m|^2$ reduces to the simple form of
\begin{equation}
\label{eq:chii}
\frac{|\chi_i|}{|\chi_m|^2}=\frac{\omega_0/\omega}{FQ}.
\end{equation}
The susceptibility loss tangent $|\chi_i/\chi_r|$ can thus be written as
\begin{equation}
\label{eq:losstan1}
\left| \frac{\chi_i}{\chi_r}\right|=\frac{(\omega_0/\omega) |\chi_r|}{FQ}\frac{|\chi_m|^2}{|\chi_r|^2}.
\end{equation}
Note that this susceptibility loss tangent is different from the permeability loss tangent $|\mu_i/\mu_r|$.  They are closely related and will be linked below.

Equation (\ref{eq:losstan1}) is exact, but a few approximations lead to simple and useful form.  Under most circumstances, the desired strong magnetic material response is achieved at frequencies around 5--10\% above or below the resonant frequency, and $\omega_0/\omega \approx 1$.  Moreover, if losses are not too severe (which is usually desired), then $|\chi_m|^2 \approx |\chi_r|^2$.  The susceptibility loss tangent is thus approximated by
\begin{equation}
\label{eq:suslt}
\left| \frac{\chi_i}{\chi_r}\right|\approx\frac{|\chi_r|}{F Q}.
\end{equation}
Provided that $|\chi_i|<|\chi_r|$, this neglected factor $\frac{\omega_0 |\chi_m|^2}{\omega |\chi_r|^2}$ is not far from unity and the approximation is accurate.  Eq. (\ref{eq:suslt}) explicitly gives the susceptibility loss tangent (from which any loss-related parameter can easily be calculated) in terms of three simple parameters: the desired real part of the susceptibility, the geometry-dependent factor $F$, and the quality factor $Q$ of an individual resonant particle.

The maximum imaginary part of the magnetic susceptibility occurs at $\omega=\omega_0$ and from (\ref{eq:chii1}) is given by $\max(\chi_i)=FQ$.  The maximum real part of the magnetic susceptibility for a Lorentzian occurs very close to the frequency where $|\chi_r|=|\chi_i|$.  Setting (\ref{eq:chii1}) equal to (\ref{eq:chir1}) and substituting the resulting relation back into (\ref{eq:chii1}), we find that
\begin{equation}
\label{eq:maxreal}
\max(\chi_r)=\frac{FQ\omega}{2\omega_0}\approx \frac{FQ}{2}.
\end{equation}
Note that this maximum $\chi_r$ is always achieved with high losses since $|\chi_r|=|\chi_i|$ at that point.

\section{Practical Implications}

As noted above, it can be shown analytically \cite{popa06srrtheory} that for a planar loop resonant particle in a cubic unit cell, $F\approx 0.16$.  This factor can be increased somewhat by narrowing the unit cell in the loop-axial direction, but a practical limit appears to be reached around $F\approx 0.5$ because the strong inter-loop interaction increases the effective inductance of a single loop and thus inhibits further increase in $F$.  This range of $F$ combines with (\ref{eq:suslt}) to give an approximate range of practically achievable susceptibility loss tangents of
\begin{equation}
\frac{2|\chi_r|}{Q}<\left| \frac{\chi_i}{\chi_r}\right|<\frac{6|\chi_r|}{Q}.
\end{equation}
The implications of this are significant.  If, for example, one wishes to design a magnetic metamaterial with a cubic unit cell which has a permeability real part of $-2$ (thus $\chi_r=-3$) at some frequency, then $|\chi_i|=|\mu_i|\approx 54/Q$ at this frequency, and the imaginary part of the susceptibility depends only on the resonant particle $Q$ factor.  A maximum $|\mu_i|$ thus immediately imposes a minimum $Q$ required of the individual resonators needed to realize this material.  If a permeability loss tangent of 0.01 is required (equivalent to $|\chi_i|=0.02$ for $\mu_r=-2$), then $Q_{\rm part}$ must be at least 2700.  This could be reduced by roughly a factor of 3 by increasing $F$ through tighter spacing of the resonant particles.  But it is unavoidable that low loss tangents require high $Q$ particles.

This analysis can be carried slightly farther to give a simple equation for permeability loss tangent.  It is straightforward to show that 
\begin{equation}
\label{eq:lteq}
|\tan \delta| = \frac{|\chi_i|}{|1+\chi_r|}\approx\frac{|\chi_r|^2}{F Q |1+\chi_r|}.
\end{equation}
The minimum value of $|\tan \delta|$ occurs at  $\chi_r=-2$ (or $\mu_r=-1$) when the above expression is valid.  Thus, provided $Q$ is not too small, the minimum achievable loss tangent is $\min(|\tan \delta|)\approx 4/FQ$.  For a cubic unit cell, $\min(|\tan \delta|)\approx 24/Q$, while for a more tightly packed lattice ($F$=0.5), $\min(|\tan \delta|)\approx 8/Q$.  $Q$ of the individual resonators thus directly constrains the minimum loss tangent of the effective medium.

The above equations are formed in terms of $Q$ because this parameter is often relatively straightforward to measure or analytically estimate.  The above analysis is also sufficiently general to apply to all resonator based metamaterials and is not limited to circuit-based elements.  For example, \cite{sarychev06} reported the theoretical design of a plasmon-based magnetic resonator for infrared metamaterials.  Provided $Q$ and $F$ can be estimated, the above equations apply equally well to such a material.

\section{Experimental Validation}\label{sec:expt}

\begin{figure}
\includegraphics[width=3.2in]{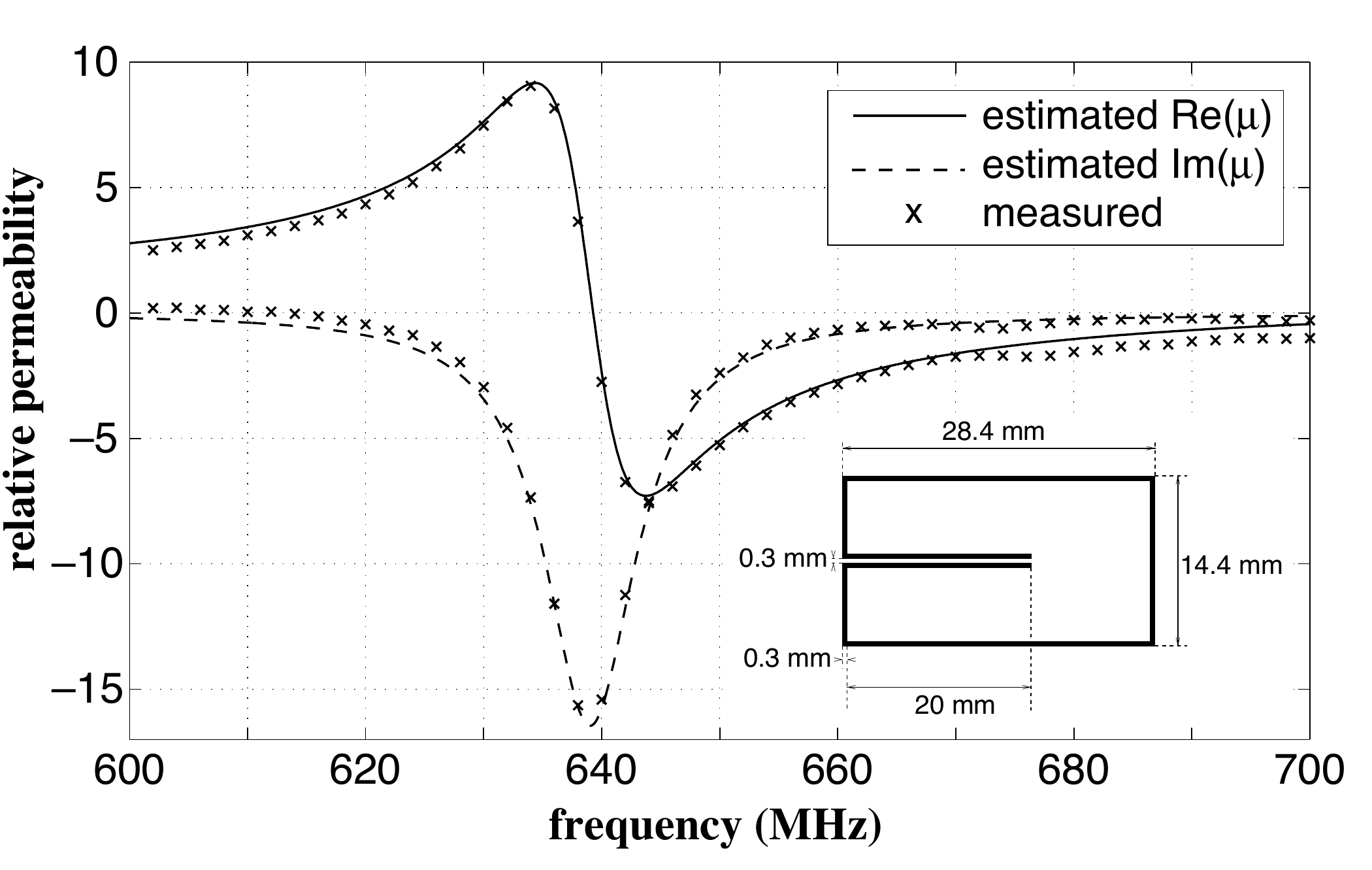}
\caption{Comparison of complex permeability of the inset loop structure measured experimentally and computed analytically from \protect{(\ref{eq:mu})}.}
\label{fig:muvalid}
\end{figure}

We validate and demonstrate this analysis by comparing the permeability measured from the $S$ parameters of a fabricated magnetic metamaterial and estimated from the above design equations using the measured $Q$ of a single resonator.  A split ring structure with dimensions shown on the inset of Figure \ref{fig:muvalid} was designed with a self resonant frequency of 685 MHz.  This copper structure was photolithographically fabricated on FR4 substrate.  The $Q$ of a single capacitively loaded loop particle was measured experimentally using an HP 8720A network analyzer by measuring $S_{11}$ of an 18 AWG wire loop placed around the particle, which exhibits a resonant response due to strong coupling of the loops when one loop is resonant.  $Q_{\rm part}=68$ was measured by fitting a Lorentzian to the difference between $S_{11}$ measured with and without the resonant particle.  $F$ for a magnetic metamaterial composed of these loops was estimated by first extracting the inductance of a single, isolated loop using Ansoft Q3D, which was found to be 101 nH.  The loop area is $A_{\rm loop}=409$ mm$^2$ and the unit cell dimensions of 15 mm by 30 mm by 16 mm ($V_{\rm cell}=7200$ mm$^2$) combine to give $F=0.29$.  This 101 nH isolated loop inductance is smaller than the effective inductance of a loop in the metamaterial array because of the mutual inductance from particle coupling.  The measured resonant frequency of the array of particles is 640 MHz, which corresponds to an increase in total inductance to 116 nH assuming the particle capacitance is constant.  This inductance correction lowers $F$ to $0.25$.  This $F=0.25$ and $Q=68$ are sufficient to compute $\mu(f)$ from (\ref{eq:mu}) or to easily estimate loss tangents from (\ref{eq:suslt}).  

A 1D array of 10 of these particles arranged in unit cells as described above was placed in the interior of a 15 cm wide microstrip waveguide designed for 50 $\Omega$ impedance.  The normal incidence $S$ parameters of the metamaterial-loaded and unloaded waveguide were measured with the same analyzer, and from these the effective permittivity and permeability of the loops were extracted \cite{smith05ret}.  Figure \ref{fig:muvalid} shows the experimentally measured complex permeability as a function of frequency.  Overlaid on the measurements is the estimated $\mu(f)$ computed using $Q=68$ and $F=0.25$ from (\ref{eq:mu}).  The agreement in both real and imaginary parts is good through the entire region of significant magnetic response, with discrepancies typically smaller than 20\%.  The measurement and estimate begin to deviate somewhat above 670 MHz, which we attribute to a combination of measurement uncertainty and weak secondary resonances due to nonuniformities in the assembly of the individual particles into the metamaterial array.

\begin{figure}
\includegraphics[width=3.2in]{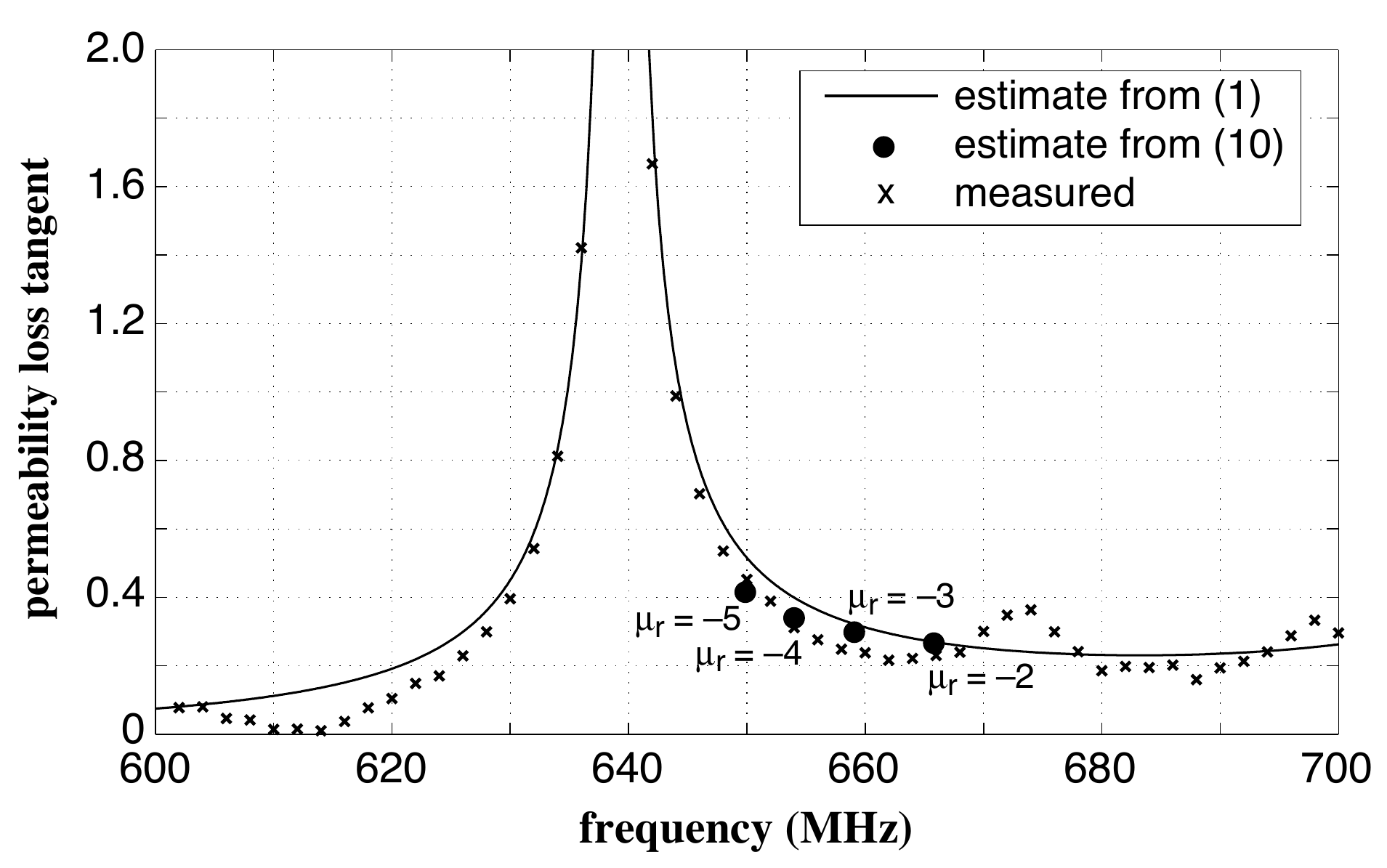}
\caption{Comparison of permability loss tangent of the fabricated loop structure measured experimentally, computed analytically from \protect{(\ref{eq:mu})}, and computed from the simple analytical approximation \protect{(\ref{eq:lteq})}.}
\label{fig:ltvalid}
\end{figure}

Figure \ref{fig:ltvalid} compares the measured and estimated permeability loss tangent $|\mu_i/\mu_r|$, which are also in good quantitative agreement.  To demonstrate the validity of the simplified approximations above, we also computed the loss tangent for several values of $\mu_r$ from (\ref{eq:lteq}), again using $Q=68$ and $F=0.25$, and overlaid them on the continuous curves.  At discrete values from $\mu_r=-2$ to $-5$, this very simple estimate is within 25\% of the measured values.  We emphasize that this accurate estimate was obtained from a simple experimental measurement of a single metamaterial particle and from a straightforward calculation of the inductance of a single particle.  The quantitative agreement is sufficiently good to enable simple and accurate estimates of the achievable loss tangent at a given value of $\mu_r$ from a measurement or simulation of $Q_{\rm part}$ and an estimate of $F$ based on the unit cell geometry.

\section{Limits on Ohmic Losses in Planar Loop Magnetic Metamaterials}

Useful limits on ohmic losses for planar loop-based magnetic metamaterials can be obtained through further analysis.  As shown above, effective material loss tangents are linearly proportional to $(FQ)^{-1}$.  Since $F=\mu_0 A_{\rm loop}^2/(V_{\rm cell}L)$ for a planar loop material \cite{popa06srrtheory}, and $Q=\omega_0 L/R$, this loss tangent parameter $(FQ)^{-1}=(V_{\rm cell}R)/(\mu_0 \omega_0 A_{\rm loop}^2)$.  The loop inductance does not play a role in the effective loss tangent of the material.  Only the unit cell size, the loop area, the loop resistance, and the resonant frequency contribute to losses.

This dependence is entirely expected because $|\chi_i|$ is proportional to the magnetostatic energy loss per unit volume per unit time.  Equation (\ref{eq:chii}) and the above show that this loss is proportional to $|\chi_m|^2 R$.  But $|\chi_m|$ is proportional to the magnetic dipole moment per unit volume, which for a current $I$ in a loop is $I A_{\rm loop}/V_{\rm cell}$.  Thus $|\chi_i|\propto I^2 R/V_{\rm cell}$, as expected because resistive losses in the loop are the only losses in this type of magnetic loop structure.

This simple expression for the loss tangent parameter $(FQ)^{-1}$ also gives important insight into how magnetic metamaterial losses scale with particle size and resonant frequency.  For a loop that comes close to spanning the side of a unit cell, the ratio $A_{\rm loop}^2/V_{\rm cell}=m l_{\rm cell}$, where $l_{\rm cell}$ is the length of a unit cell in the propagation direction and $m$ is the number of loops per cubic volume.  For example, $m=2$ applies to a unit cell that is half as wide as it is long so that there are two loops in a cubic volume.  Using the above expression for $(FQ)^{-1}$, the susceptibility loss tangent is closely approximated by 
\begin{equation}
\left|\frac{\chi_i}{\chi_r}\right|\approx\frac{|\chi_r|R}{m l_{\rm cell} \omega_0\mu_0}=\frac{|\chi_r|}{2\pi m} \frac{R}{Z_0}\frac{\lambda_0}{l_{\rm cell}},\label{eq:losswithr}
\end{equation}
where $Z_0=377$ $\Omega$ is the impedance of free space and $\lambda_0$ is the free space wavelength at the particle resonant frequency $\omega_0$.  From (\ref{eq:lteq}) and the analysis immediately following, we can also write the minimum permeability loss tangent as
\begin{equation}
\min(|\tan \delta|) \approx \frac{2}{\pi m} \frac{R}{Z_0}\frac{\lambda_0}{l_{\rm cell}},\label{eq:ltwithr}.
\end{equation}
The size of the unit cell relative to wavelength plays a role as important as $R$ in the metamaterial losses---the larger the unit cell compared to wavelength, the lower the loss.  This indicates that simply scaling down the size of a loop particle in order to better approximate a continuous medium will increase the losses if $R$ remains constant.  And simple scaling that decreases the loop length and loop width by the same factor will keep $R$ per particle close to constant.  This suggests controlling losses may be a challenge in magnetic metamaterials in which the resonant particles are very small compared to wavelength.  Increasing $m$ reduces the losses the losses, but there is a practical limit to how many loops can be packed into a cubic volume.  

We can go even further by noting that the resistance of planar loops is constrained based on their geometry.  We assume that the metal used for the loops is room temperature copper.  If we further assume that the thickness of the loop copper is large compared to the skin depth, which will minimize the loop resistance, then $R$ is simply a function of the number of individual geometric squares that compose the loop.  The surface resistance of a thick (compared to skin depth) conducting sheet is $(\sigma\delta)^{-1}$ ohms per square, where $\sigma$ is the conductivity of the metal and $\delta$ is the skin depth.  For copper, $\delta(f)= 2.1/\sqrt{f({\rm GHz})}$ $\mu$m and thus the surface resistance of the trace is $R_{\rm s}=0.0082\sqrt{f({\rm GHz})}$ ohms per square.  Thus, a loop that is thick compared to skin depth and composed of $n_{\rm sq}$ squares of copper results in
\begin{equation}
{\rm min} |\tan \delta| \approx 1.4\times10^{-5} \frac{n_{\rm sq}}{m} \frac{\lambda_0}{l_{\rm cell}} \sqrt{f({\rm GHz})}.
\label{eq:numlt}
\end{equation}
This estimate should be considered as an approximate lower bound as many favorable assumptions were made in its derivation.  For example, the metal traces are assumed to be at least several skin depths thick, and thinner traces will increase losses.  The current was also assumed to flow uniformly across the width of the trace, and proximity effects can break this uniformity and increase losses.  We also emphasize that this loss accounts only for ohmic losses in the copper.  Dielectric losses in the necessary capacitance will increase the total loss.  The weak frequency dependence of this loss bound originates in the frequency dependence of skin depth.  

This expression enables the easy computation of a realistic lower bound on the effective loss of a magnetic metamaterial due to ohmic losses.  For example, let us consider a square loop composed of 12 squares at 1.5 GHz.  Twelve squares is a small but practical number that results in a copper width of 25\% of the size of the loop from outer edge to outer edge.  At 1.5 GHz, the resistance of this loop is no less than $0.10 \times \sqrt{1.5}$=0.123 ohms.  For a cubic unit cell ($m=1$) and ${\lambda_0}/{l_{\rm cell}}=5$ (essentially the minimum for which the loops can still be have like an effective material),  the minimum achievable permeability loss tangent for planar loops of room temperature copper is ${\rm min} |\tan \delta| \approx 0.001$ according to (\ref{eq:numlt}).  This can be reduced by packing more loops into a cubic cell volume, but practical considerations like substrate thickness may make it difficult to achieve an $m$ greater than 4 or 5.

This expression can also be applied to the fabricated and measured particle as described in Section \ref{sec:expt}.  This is a thin loop with $n_{\rm sq} \approx 285$ according to the dimensions in Figure \ref{fig:muvalid}.  The dimensions also give $A_{\rm loop}^2/V_{\rm cell}=m l_{\rm cell}=23.2$ mm, or $\lambda_0/m  l_{\rm cell}=20.2$ at the 640 MHz resonant frequency.  These numbers and (\ref{eq:numlt}) yield ${\rm min} |\tan \delta| \approx 0.065$.  This is approximately 3 times smaller than the measured minimum permeability loss tangent (consistent with it being a lower bound).  The dielectric losses from the FR4 substrate in between the capacitor traces are not included in this analysis and probably produce the difference.  Nevertheless, this simple analytical estimate produces a lower bound not too much lower than what is measured, confirming that (\ref{eq:numlt}) represents a strong lower bound that is simple to calculate analytically.  We emphasize that the estimate based on the measured $Q$ of a single particle automatically include both conductive and dielectric losses, and thus result in a much more accurate loss tangent estimate.

%This analysis shows clearly that magnetic metamaterial permability loss tangents on the order of 0.1 are easy to achieve, that those on the order of 0.01 are challenging but potentially feasible, and that those on the order of 0.001 are probably not feasible for planar loop materials.  These limits are a sublinear function of frequency and thus this general guideline applies reasonably well between 100 MHz and 10 GHz.  Aside from a layout that employs as much copper as possible to reduce the loop resistance, there are two key design parameters that control loss.  One is the number of loops per cubic volume, and the other is the size of the unit cell compared to wavelength.  Increasing the number of loops per cubic volume decreases losses because it reduces the amount of current, and therefore ohmic losses, per loop in order to create a fixed magnetic dipole moment per unit volume.  Increasing the number of unit cells per wavelength also decreases losses because larger loops are able to create a larger magnetic dipole moment per unit of ohmic loss.

%For square loops with side length $l$ typical for magnetic metamaterial designs, we find empirically that $L\approx 3\mu_0 l$.  A good example is the loop shown in Figure \ref{fig:qsim} with a side length of 30 mm and a 95.7 nH inductance that is close to $3\mu_0 l=113$ nH.  This means that $l_{\rm eff}\approx 3 l_{\rm cell}$ and 

%still need more careful scaling analysis and need to hit the spiral vs. loop and the various scaling arguments.

\section{Conclusions}

In summary, we showed through full wave numerical simulations that $Q_{\rm mat}$ of the overall effective material resonant response is essentially equal to $Q_{\rm part}$ of the individual resonant particles that make up the metamaterial.  We derived simple equations for the effective properties of resonant metamaterials in terms of this quality factor $Q$ of an individual resonator.  These relations enable simple calculations of the $Q$ required to achieve desired resonant metamaterial effective parameters, thereby enabling realistic and accurate analytical predictions of material properties simply from $Q_{\rm part}$, which often can analytically estimated or measured from small material samples or even individual particles.  We experimentally demonstrated this approach by measuring $Q_{\rm part}$ of a single fabricated metamaterial particle.  This single particle measurement was used to semi-analytically predict the effective permeability of a metamaterial slab which was was within 20\% and better agreement with the permeability extracted from a much more complicated, full $S$ parameter measurement of the slab.  Although derived and validated specifically for loop-based magnetic metamaterials, the general design equations apply equally well to any metamaterial, electric or magnetic, based on resonant particles.

These design equations were analyzed further to derive simple, numerical lower bounds on the ohmic losses in planar magnetic metamaterials.  It is shown that the effective magnetic loss tangent of the such a material depends on the resistance of the loop, the linear size of the loop relative to wavelength, and the number of loops per cubic volume in the metamaterial.  The resulting expression is a lower bound on losses in magnetic metamaterials that is simple to calculate analytically and shows that minimum achievable loss tangents are close to 10$^{-3}$ at frequencies between roughly 100 MHz and 10 GHz, and that achieving this bound requires thick copper traces for a loop only a few times smaller than a wavelength.  Because losses vary inversely with the size of the particle relative to a wavelength for a planar loop magnetic metamaterial, realizing low losses in such a material with electrically very small particles may not be possible.

%Simplifies the design process.

%although done for magnetic loop material, can easily apply to other resonant metamaterials.  only difference would be in computing $F$.

%\section{Validation}
%
%In HFSS, use a discrete resistor so loss is known, compute Q analytically, show that it is the same as Qmeas.  Then use our expressions to compute real and imaginary part of susceptibility.
%
%also validate using loop capacitor medium.  Measure Q, then do a retrieval on the collection.
%
%
%
%Probably ought to do a second paper on scaling and loss estimation.  compare E particles and H particles here.

%\clearpage

\bibliographystyle{IEEEtran}
%\bibliography{cummer,cloaking,metamat,numericalem,books}% Produces the bibliography via BibTeX.

\end{document}